# Anomalous Fiber Optic Gyroscope Signals Observed above Spinning Rings at Low Temperature

**M Tajmar, F Plesescu and B Seifert**

Space Propulsion & Advanced Concepts,
Austrian Research Centers GmbH – ARC, A-2444 Seibersdorf, Austria

E-mail: martin.tajmar@arcs.ac.at

**Abstract**. Precision fiber optic gyroscopes were mounted mechanically de-coupled above spinning rings inside a cryostat. Below a critical temperature (typically <30 K), the gyroscopes measure a significant deviation from their usual offset due to Earth's rotation. This deviation is proportional to the applied angular ring velocity with maximum signals towards lower temperatures. The anomalous gyroscope signal is about 8 orders of magnitude smaller then the applied angular ring velocity, compensating about one third of the Earth rotation offset at an angular top speed of 420 rad/s. Moreover, our data shows a parity violation as the effect appears to be dominant for rotation against the Earth's spin. No systematic effect was found to explain this effect including the magnetic environment, vibration and helium gas friction suggesting that our observation is a new low temperature phenomenon. Tests in various configurations suggest that the rotating low temperature helium may be the source of our anomalous signals.

## 1. Introduction

Frame dragging is a phenomenon in general relativity that causes spinning matter to drag space-time in its vicinity. According to Einstein's theory, this effect is so weak that it required astronomical observations and precision tests with satellites to detect it [1]. Apart from such a weak classical gravitational spin-coupling [2], a number of laboratory experiments have been carried out in the past to look for new macroscopic interactions coupling mass to intrinsic spin mostly using torsion balances and polarized spin-sources [3-4]. For example, Moody and Wilczek [5] proposed a monopole-dipole interaction coupling electron spin to an unpolarized nucleon mediated by the axion boson violating both parity and time reversal. No such interaction could be found up to now and experimental limits on possible electron spin interactions could be placed for an interaction range between 5 mm and 1 m.

Recently, Tajmar and de Matos [6-8] predicted that spinning superconductors or superfluids instead of electron spin polarized materials might produce much larger non-classical frame-dragging fields in order to explain a reported Cooper-pair mass anomaly in niobium [9-10]. Other theoretical concepts were proposed supporting this conjecture [11-13]. An initial experiment was setup to investigate frame-dragging at low temperatures using accelerometers that indicated first positive results [14-15]. In order to suppress large vibration-dependent systematic effects and to increase the measurement resolution, the accelerometers were supplemented with optical gyroscopes. Here too we found an anomalous gyro response above spinning rings at temperatures below 30 K which was increasing towards lower temperatures [16]. The effect was found not to depend on superconductivity and it showed a parity violation such that the signal was dominant for clockwise rotation only. For a niobium ring at 4 K, a coupling factor of $3.2\pm0.5\times10^{-8}$ for clockwise rotation between gyro signal and applied angular ring velocity was measured 4 cm above the ring [16]. Graham et al [17] performed a test to look for frame-dragging from a spinning lead disc at 4 K using the world's largest ring-laser gyro and failed to detect a signal within $1.7\pm3.8\times10^{-7}$ assuming a dipolar distribution, a sensitivity which is about two orders of magnitude worse compared to our design (their experimental values were re-assessed in [16]). Moulthrop [18] investigated frame-dragging-like signals from rotating superfluid helium and put an upper limit coupling-factor of 0.05 which is 7 orders of magnitude worse compared to our sensitivity. We recently implemented a military-grade fiber-optic-gyroscope (Optolink SRS-1000) and varied the experimental configuration in order to isolate the cause for the anomalous signal boosting our sensitivity again by 2-3 orders of magnitude. This paper will present an overview of our gyroscope experiments to test frame-dragging or spin-coupling forces at low temperatures.

## 2. Experimental Results

*2.1. Configuration*
The core of our setup is a rotating ring inside a large cryostat. Several rings have been used so far including niobium, an aluminum alloy (AlCuMgPb containing >90% Al, we will refer to this alloy as Al through the rest of the text), stainless steel 304 as well as YBCO and TEFLON to test classical low-temperature and high-temperature superconductors as well as non-superconducting materials. The Nb, Al, and stainless steel rings had an outer diameter of 150 mm, a wall thickness of 6 mm and a height of 15 mm, whereas the YBCO and TEFON had a slightly larger outer diameter of 160 mm. If ring and sample holder are not of the same material (see table 1), then the ring is glued inside an Al sample holder using STYCAST cryogenic epoxy. The bottom plate of the sample holder is made out of stainless steel as well as the rotating shaft. In general, only non-magnetic materials were used throughout the facility. The temperature of the ring is monitored with a silicon diode (Lakeshore DT-670B-SD) which is connected to a miniature collector ring. A field coil is used to trap magnetic fields with the superconducting rings during cool down to verify the correct temperature read-out since the magnetic field will be lost after passing the material's critical temperature again which is a known reference point. All experimental data is summarized in table 1.

**Table 1.** Comparison of All Experimental Data at T=4-6 K at the Gyro Location.

| Setup | Ring Material | Sample Holder Material | Type | Coupling Factor (Gyro Output/$\omega$) × $10^8$ | |
|---|---|---|---|---|---|
| | | | | CW | CCW |
| A | Nb | Al | Ring | 5.7 ± 0.4 | 4.8 ± 0.5 |
| | YBCO | Al | Ring | 3.1 ± 0.4 | 0.3 ± 0.4 |
| | TEFLON | TEFLON | Ring | 3.4 ± 0.2 | -0.5 ± 0.2 |
| | - | - | No Sample | -0.1 ± 0.2 | -0.3 ± 0.1 |
| | | | | | |
| B | Stainless Steel | Stainless Steel | Ring | 3.4 ± 0.5 | -4.7 ± 0.9 |
| | Al | Al | Ring | 2.1 ± 0.8 | -2.2 ± 0.5 |
| | Al | - | Disc | 0.3 ± 0.1 | -0.3 ± 0.1 |
| | | | | | |
| C | Al | Al | Cup with Fins | 0.17 ± 0.03 | 0.01 ± 0.09 |
| | Al | Al | Ring | -0.03 ± 0.03 | -0.04 ± 0.04 |
| | Nb | Al | Ring | -0.12 ± 0.05 | -0.12 ± 0.08 |
| $LN_2$ (77-90 K) | YBCO | Al | Ring | 0.0 ± 0.01 | 0.01 ± 0.01 |
| | | | | | |
| Graham et al [17] | Pb | - | Disc | -5.3 ± 8.5 | 37.7 ± 13.2 |

The gyroscopes are mounted together with accelerometers (Colibrys SiFlex1500 and SiliconDesigns 1221), magnetic field (Honeywell SS495A1) and temperature sensors usually a few centimeters away from the ring at room temperature. The different configurations are illustrated in figure 1 and are outlined in the following chapters.

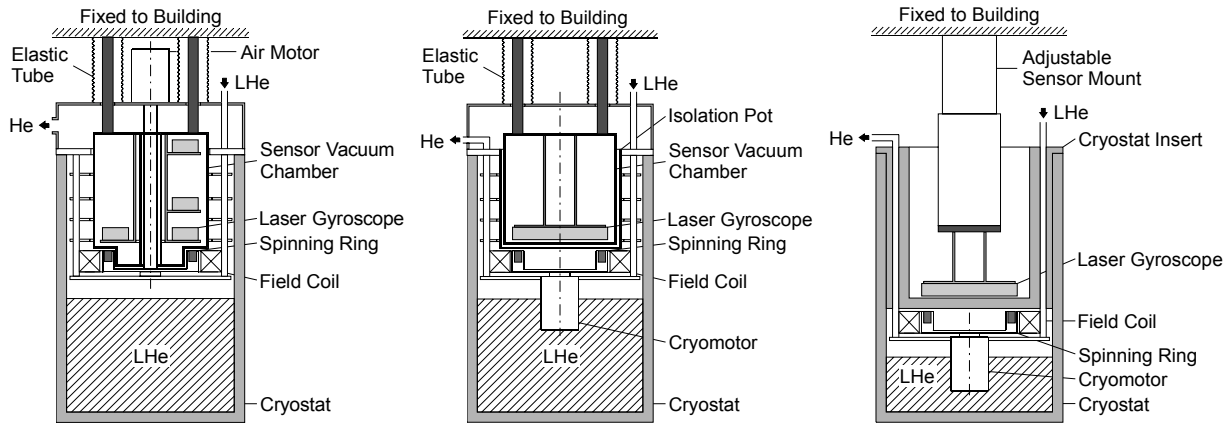

**Figure 1.** Setup A, B, and C (from Left to Right).

*2.2. Setup A*
Here, the rings were rotated using a pneumatic air motor (Düsterloh PMW 400 Z24) to minimize any electromagnetic influence. The motor and the ring assembly are mounted on top and inside a liquid helium cryostat respectively, which is stabilized in a 1.5 t box of sand to damp mechanical vibrations induced from the spinning ring. The sensors are mounted inside an evacuated chamber made out of stainless steel, which acts as a Faraday cage and is directly connected by three solid shafts to a large structure made out of steel that is fixed to the building floor and the ceiling. The sensors inside this chamber are thermally isolated from the cryogenic environment due to the evacuation of the sensor chamber and additional MLI isolation covering the inner chamber walls. Heaters and temperature sensors along the support structure inside the sensor chamber maintain a room temperature environment to obtain a good sensor temperature drift stability. Only flexible tubes along the shafts and electric wires from the sensor chamber to the upper flange establish a very weak mechanical link between the sensor chamber and the cryostat. More details on the experimental setup can be found in [14-16]. A minimum distance of 5 mm between the sensor chamber and all cryostat parts such as the spinning ring and motor shaft is kept throughout the assembly. The difference to our earlier work was that previously the gyros were mounted on three levels using two rods which were replaced by a tubular support in order to increase the stiffness of the gyro support.

We used a KVH DSP-3000 fiber-optic gyro with a Random-Angle-Walk (RAW) of $2 \times 10^{-5}$ rad/s and a resolution of $1 \times 10^{-6}$ rad/s. Laser gyros are sensitive to magnetic fields due to the Faraday effect. Using µ-metal shielding, we reduced the magnetic sensitivity along the gyro's rotation axis to 0.04 rad/s/T. During rotation, we measured maximum magnetic fields in the order of hundreds of nT which corresponds to magnetic field offsets below the gyro's resolution. Four gyros are mounted in three positions (reference, middle and above ring) where two of them are mounted above the rotating ring (one with gyro axis up and one down to investigate any offset issues, see figure 1). The axial distance from the top ring surface to the middle of each gyro position is 45.8 mm, 92.8 mm and 220.8 mm respectively. All sensors are mounted at a radial distance of 53.75 mm.

The following data was obtained using the pneumatic air motor following a standardized speed profile which lasts about 20 seconds. The profile is subdivided into 5 sectors with pre-defined time intervals: rest, acceleration, maximum speed, de-acceleration and rest. Each profile is repeated first in clockwise direction (spin vector of ring points downwards) and then in counter-clockwise direction. Initially, liquid helium is filled into the cryostat until it reaches the bottom plate of the sample holder. Then the speed profiles are repeated usually up to 80 times until liquid helium is filled again into the cryostat. During operation of the facility, the temperature of the ring is slowly rising acquiring data at different temperatures (temperature stability at gyro location is better than 0.02°C over one profile).

Figure 2 shows the reaction measured by the gyroscope during the rotation of the YBCO ring within a temperature interval of 4-15 Kelvin, where the positive angular velocity indicates clockwise rotation (Earth offset subtracted). No anomalous signal was measured when the sample holder was removed but the rest of the assembly (motor, shaft, bearings) remained in place [16]. Random noise was reduced by applying a 20 pt digital moving average (DMA) filter to each gyro and encoder velocity output. A sampling rate of 13 Hz ensured that the averaging windows is always smaller than the signal peak during the profile. The plots were obtained by signal averaging of typically 200 profiles for each sample whose average individual profile temperature falls within the 4-15 Kelvin temperature range.

Figure 3 shows the variation of the average from the two above ring gyros output versus the ring's angular velocity in clockwise orientation over a temperature range from 4-55 Kelvin in 2 Kelvin intervals for the Nb, YBCO and TEFLON samples. First, the mean value and standard deviation of the gyros and the angular velocity of the two zero and maximum speed sectors are calculated using averaged and filtered profiles for a defined 2 Kelvin temperature interval as described above with typically 20 profiles per interval for statistical confidence. Then a least square linear regression between the three data points and their respective standard deviation is performed. The linear fit uncertainty is plotted as the coupling factor sigma in figure 3. The mean value and standard deviation of the averaged profiles temperatures per interval is plotted along the temperature axis.

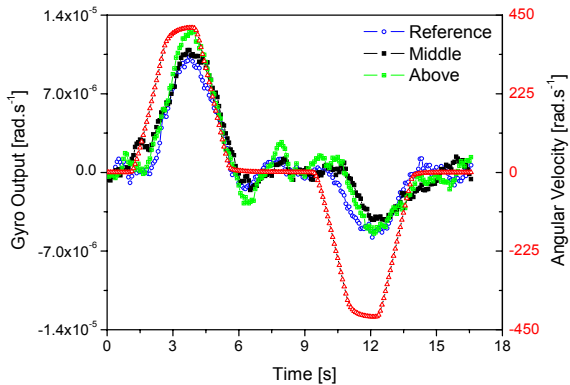

**Figure 2.** KVH Laser Gyro Output for YBCO at Different Positions Versus Applied Angular Velocity (Δ) between a Temperature of 4-15 Kelvin – Setup A.

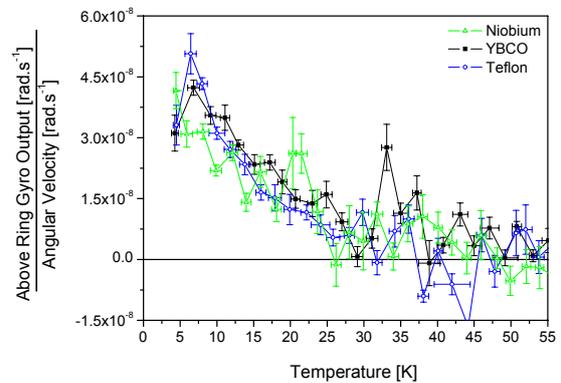

**Figure 3.** Variation of Above Ring Gyro Output versus Angular Velocity with Temperature (for Clockwise Rotation) – Setup A.

The results show that the rotation of our sample rings can indeed be seen by rigidly mounted gyroscopes at a distance once the sample ring passed a critical temperature below 30 K. At 4 K and a top speed of 420 rad/s, the anomalous gyro signal is as large as one third or the Earth's rotation. Their behaviour can be summarized as follows:

First, the gyro outputs show a parity violation between clockwise and counter-clockwise rotation (the clockwise signal is larger than the counter-clockwise signal). It is important to note that for the case of the above-ring gyros, if the gyro's orientation is flipped, the effect sign flips as well and the large effect is only present for the same clockwise-rotating of the spinning ring. This measurement tells us that the effect seems to be rotationally symmetric and that the source for the parity anomaly does not result from the gyro itself.

Second, there is only a weak dependence on the sample ring material (within a factor of two along the temperature). The effect appears below 25-30 K. Previously [16], Al and Nb samples showed an effect below 15-20 K, here Nb is shifted towards a critical temperature of 25 K.

And third, the signal does not decay from the above to the reference position as one would expect from a dipolar-field distribution. Systematic effects were analysed in detail such as magnetic field influence, gas drag, sensor location and vibration [16]. For example, the gyro would need to tilt by 20° in order to explain our results as a change in the offset due to the Earth's rotation. Since the magnetic field is constantly monitored and the contribution to the gyro signal is below its resolution, only gas drag acting on the sensor chamber and vibration could in principle contribute to our measurement. Although the analysis showed that gas dragging torques should be two orders of magnitude below our sensitivity and that the gyros are not affected by vibration in our environment (monitored with the accelerometers), they remained the error source with the largest uncertainties.

*2.3. Setup B*

In order to minimize the helium gas contact with the sensor chamber and associated unknown systematics from Setup A, we implemented an isolation pot between the spinning ring and the sensor chamber. A 5 mm minimum distance between the sensor chamber and the isolation pot is being kept. Only gas leaks at the cryostat's top flange allow the upper part with the flexible tubes to be slowly filled with helium in order to avoid ice formation along the sensor chamber walls. In order to further improve the resolution, we implemented an Optolink SRS-1000 fiber-optic gyroscope only 16 mm above the spinning ring and a cryomotor (Phytron VSH-UHVC 100) as shown in figure 1. The SRS gyro has a RAW of $3\times10^{-7}$ rad/s and a digital resolution of $6\times10^{-12}$ rad/s with a superior temperature stability and even further reduced magnetic sensitivity compared to the KVH gyro. Using again μ-metal shielding, the maximum magnetic field during rotation was below 1 μT. The magnetic sensitivity along the gyro's rotation axis is $4.8\times10^{-3}$ rad/s/T which corresponds to a maximum magnetic field offset of $4.8\times10^{-9}$ rad/s. The cryomotor allows only speeds up to 100 rad/s with a maximum angular acceleration of 33 rad/s$^2$, however, the top speed duration was increased from 2 s up to 30 s. This longer duration allows to better investigate the relation between anomalous signal and applied angular velocity as well as better averaging.

The results shown in figures 4 are similar to the analysis used for Setup A but with a 50 pt DMA filter. Stainless steel and aluminum rings produced similar effects compared to Setup A, however, here clockwise and counter-clockwise rotations show signals in the same direction. The signal strengths are 3 orders of magnitude above the maximum magnetic field offset. Helium gas drag or helium flow induced thermal effects (isolation pot) and vibration effects (the cryomotor noise is negligible compared to the pneumatic air motor) are greatly suppressed in comparison to Setup A. Due to the high resolution of the SRS gyro, we could also evaluate the scaling of the effect with the ring angular velocity by slowly ramping up to full speed in 30 s and going slowly back to zero speed again in 30 s. By performing a least square linear regression analysis between angular velocity and gyro output, we could calculate a correlation factor of 0.91.

Previous modelling of the effect [16] seemed to agree well with the assumption that the effect can be best described as an amplification mechanism to classical frame-dragging. In order to amplify the effect we therefore spun a rotating aluminum disc with the same outer diameter instead of the usual ring since the effect was believed to scale with the angular momentum of the spinning sample.

The result in figure 4 however showed that the effect in this case nearly vanished. Figure 5 shows the same plot enhanced with a 200 pt DMA filter. It is clear that the effect is still there but reduced by about an order of magnitude.

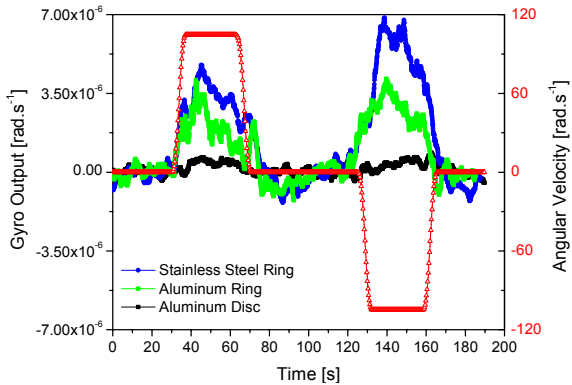

**Figure 4.** SRS Laser Gyro Output for Stainless Steel and Aluminum Versus Applied Angular Velocity (Δ) at a Temperature of 4 Kelvin – Setup B.

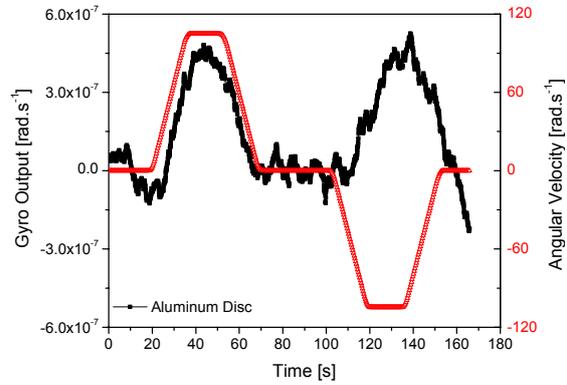

**Figure 5.** High Resolution SRS Laser Gyro Output for Aluminum Disc Versus Applied Angular Velocity (Δ) at a Temperature of 4 Kelvin – Setup B

Comparing the coupling factors of the spinning rings in figure 6 and 7 with the ones from Setup A in figure 3, the signal strengths are very similar although two different gyros and configurations were used. Only a couple of profiles were averaged over each temperature interval to obtain the plots contrary to the several tens of profiles needed for the KVH gyro due to the better resolution of the SRS gyro. More data runs are necessary to better evaluate the signal variation over the temperature range. However, it is clear that the effect starts in a temperature range between 10-15 K, which is lower compared to Setup A.

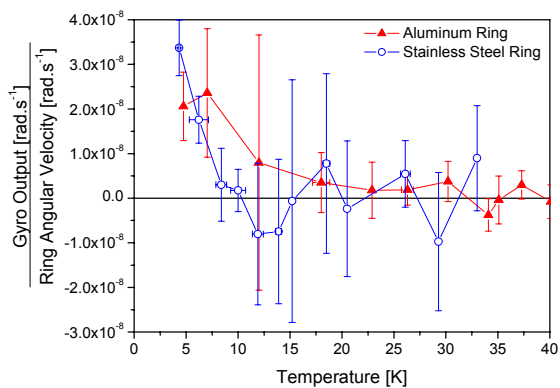

**Figure 6.** Variation of SRS Gyro Output versus Angular Velocity with Temperature (for Clockwise Rotation) – Setup B.

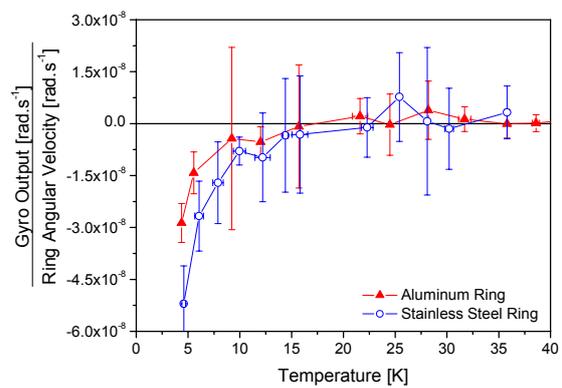

**Figure 7.** Variation of SRS Gyro Output versus Angular Velocity with Temperature (for Counter-Clockwise Rotation) – Setup B.

From Setup A and B, we learned that the maximum anomalous gyro signals appeared when we have a ring spinning with maximum signals close to 4 K. The removal of the ring in Setup A as well as the replacement of the usual rings with an aluminium disc in Setup B removed the effect by at least an order of magnitude. Therefore our best guess is that the origin of the effect is connected to the low temperature helium which did then not spin well without the ring in Setup A and was displaced when the aluminium disc occupied its space in Setup B. This would also explain why the material choice between the rings only showed a weak influence on the observed phenomenon that could be well attributed e.g. to a different surface roughness between the rings. It may also explain the reason why the above, middle and reference gyros in Setup A always measured the same anomalous signal since the helium gas was allowed to flow along the sides of the sensor chamber which may have carried the effect with it. Especially due to the isolation pot, there are now no systematic effects that we are aware of that could explain the results obtained with Setup B.

### 2.4. Setup C

In order to completely isolate the spinning ring from the gyro and to minimize helium gas flow along the side walls, we built a cryostat insert (evacuated and MLI isolated) into our original cryostat. The SRS gyro is now truly mounted outside the cryostat facility and apart from the well monitored magnetic influences, no facility-induced artefacts are known. A major difference with respect to the previous setups is that the helium gas is only allowed to flow upwards in a tube, there is no room for the helium to expand along the side walls along the height of the gyro. If our hypothesis that the rotating helium is the main source of the anomalous gyro effect is correct, then this should significantly effect our measurement. The gyro distance to the ring needed to be enlarged to 38.5 mm which is more than double compared to Setup B. Now the gyroscope has a distance of 5 mm to the bottom of the cryostat. The cryomotor assembly was moved towards the bottom of the cryostat to minimize the liquid helium consumption.

A first measurement was done using the usual Al ring from the previous setups which showed a zero result. The fact that the Al ring effect was at least reduced by nearly two orders of magnitude between Setup B and C clearly demonstrates that the origin of the anomalous signals is related to the helium flow. Then the Nb ring was used and again a zero result within 2σ from the measurement accuracy was obtained (see table 1). In order to test our helium hypothesis, we manufactured a thin-wall cup with two fins (2 crossed fins to effectively spin the liquid helium) out of Al which we mounted instead of the usual rings. We then filled the cryostat up to the top with liquid helium in order to fill the cup as well. The results are shown in figure 8 and 9 using again a 200 pt DMA filter. The noise especially in the counter-clockwise data may be again explained due to the fewer number of profile samples used which we will improve in future measurements. Nevertheless, we again obtained an anomalous gyro response within a 5σ measurement in the clockwise direction at 6 K.

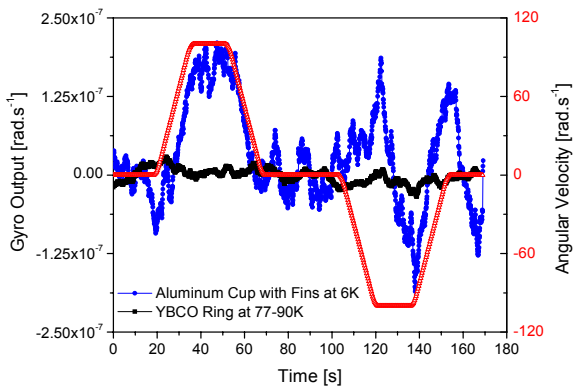
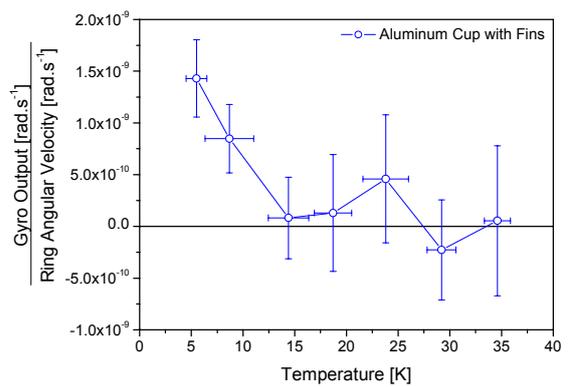

**Figure 8.** High Resolution SRS Laser Gyro Output for Aluminum Propeller and YBCO Ring Applied Angular Velocity (Δ) – Setup C.

**Figure 9.** Variation of SRS Gyro Output versus Angular Velocity with Temperature (for Clockwise Rotation) – Setup C.

The shape of the effect is similar to our Setup A data. The signal strength however is now reduced by about a factor of 25, which is still two orders of magnitude above our maximum magnetic field influence. A reduction was already expected due to the larger gyro-ring distance in Setup C and the fact that the helium gas can not flow along the sensor chamber side walls. Hopefully we should now be able to map the effect dependence along the spinning axis via the adjustable gyro mount in the future. Using the YBCO ring and liquid nitrogen gives a null reference as shown in figure 8.

We also mounted 10 accelerometers on top of the gyroscope in tangential, radial and vertical direction at various radial locations in order to characterize the acceleration environment in a facility-vibration-free environment (our dominant error source in the first accelerometer measurements [14-15]). We can put a limit on acceleration fields in all three directions to ±30 μg for all types of rings used (Nb, Al and YBCO) at 4 K and 77 K for YBCO with ring angular top speeds of 100 rad/s and accelerations of 33 rad/s$^2$. Also magnetic fields generated by the field coil of 100 μT and 1 mT during the test runs did not show a change in the acceleration signals. Due to the much smaller angular acceleration compared to our earlier air motor measurements ($\cong$1500 rad/s$^2$), we were not able to confirm or refute our previous accelerometer measurements [14-15] without the vibration error source. However, our results put strong limits on all proposed theories suggesting frame-dragging or vertical gravitational fields generated from spinning superconductors [11-14].

## 3. Conclusions
Anomalous signals from two different fiber-optic gyroscopes (KVH DSP-3000 and Optolink SRS-1000) were observed above spinning rings at temperatures below 30 K. Results from different configurations suggest that the origin is probably connected to the rotating helium and not the angular momentum of the spinning samples as it was suspected from earlier measurements. Our observed signal strengths are not ruled out by any other experiment up to our knowledge [17-18] and systematic effects appear to be at least two orders of magnitude below all reported measurements for the various setups. The gyro signal seems to follow the rotating ring velocity with high correlation. Compared to classical frame-dragging spin-coupling predictions, our signals are up to 18 orders of magnitude larger. This suggests that the observed phenomenon is new and without explanation so far.

**Acknowledgments**
This work was funded by the Austrian Research Centers. We would like to thank N. Buldrini, the Institute for Advanced Studies in Austin, G. Hathaway and M. Millis for useful discussions.